\documentclass{midl} 

\usepackage{graphicx} 

\usepackage{siunitx} 
\DeclareSIUnit\px{px}
\DeclareSIUnit\mucron{\upmu m} 
\usepackage{upgreek} 
\usepackage[format=plain]{caption} 
\let\clearpage\relax 

\jmlrproceedings{MIDL}{Medical Imaging with Deep Learning}
\jmlrpages{}
\jmlryear{2021}

\jmlrworkshop{Short Paper -- MIDL 2022 submission}
\jmlrvolume{-- Under Review}
\editors{Under Review for MIDL 2022}

\title[Deriving CK expression from DAPI channel via \textsc{dapi2ck}]{Novel Deep Learning Approach to Derive Cytokeratin Expression and Epithelium Segmentation from DAPI}

\midlauthor{\Name{Felix Jakob Segerer}\footnotemark[1]\nametag{$^{1}$} \Email{felix.segerer@astrazeneca.com}\\
\Name{Katharina Nekolla}\footnotemark[1]\nametag{$^{1}$} \Email{katharina.nekolla@astrazeneca.com}\\
\Name{Lorenz Rognoni}\footnotemark[2]\nametag{$^{1}$} \Email{lorenz.rognoni@ultivue.com}\\
\Name{Ansh Kapil}\nametag{$^{1}$} \Email{ansh.kapil@astrazeneca.com}\\
\Name{Markus Schick}\nametag{$^{1}$} \Email{markus.schick@astrazeneca.com}\\
\Name{Helen Angell}\nametag{$^{2}$} \Email{helen.angell@astrazeneca.com}\\
\Name{Günter Schmidt}\nametag{$^{1}$} \Email{guenter.schmidt@astrazeneca.com}\\
\addr $^{1}$ AstraZeneca Computational Pathology, Oncology R\&D, Munich, Germany \\
\addr $^{2}$ AstraZeneca Translational Medicine, Oncology R\&D, Cambridge, UK
\footnotetext[1]{Contributed equally}
\footnotetext[2]{Was working for AstraZeneca Computational Pathology by the time this study was conducted. Currently working for: \textit{Ultivue Europe, Segrate Milan, Italy}}
}

\begin{document}

\maketitle

\begin{abstract}
Generative Adversarial Networks (GANs) are state of the art for image synthesis. Here, we present \textsc{dapi2ck}, a novel GAN-based approach to synthesize cytokeratin (CK) staining from immunofluorescent (IF) DAPI staining of nuclei in non-small cell lung cancer (NSCLC) images. We use the synthetic CK to segment epithelial regions, which, compared to expert annotations, yield equally good results as segmentation on stained CK. Considering the limited number of markers in a multiplexed IF (mIF) panel, our approach allows to replace CK by another marker addressing the complexity of the tumor micro-environment (TME) to facilitate patient selection for immunotherapies. In contrast to stained CK, \textsc{dapi2ck} does not suffer from issues like unspecific CK staining or loss of tumoral CK expression.
\end{abstract}


\section{Introduction}

Understanding the TME is key to understand the mechanisms behind immunotherapy of cancer. The complexity of the TME can be assessed via mIF as it enables the spatially resolved evaluation of multiple biomarkers on the same tissue section. However, the number of markers within a panel is limited. Deep learning can assist here by leveraging information from one marker to predict expression of other markers, hence allowing to replace such markers by complementary ones. 
Based on GANs, which are state of the art for such data synthesis tasks, \citet{Isola2016} proposed a powerful framework for supervised image-to-image translation called pix2pix. In the field of digital pathology, a similar approach was used by \citet{Burlingame2018} to predict tumor markers including CK from hematoxylin and eosin stained images. Here, we trained a pix2pix to predict CK expression by leveraging the morphological information of DAPI stained nuclei in NSCLC tissue sections.

\begin{figure}[htbp]
\floatconts
  {fig_1}
  {\caption{Two-step approach to segment epithelium from DAPI: 1. The generator of the \textsc{dapi2ck} network is used to generate a synthetic CK staining from DAPI. 2. The synthetic CK channel serves as input for a model segmenting epithelial regions.}}
  {\includegraphics[width=0.75\linewidth]{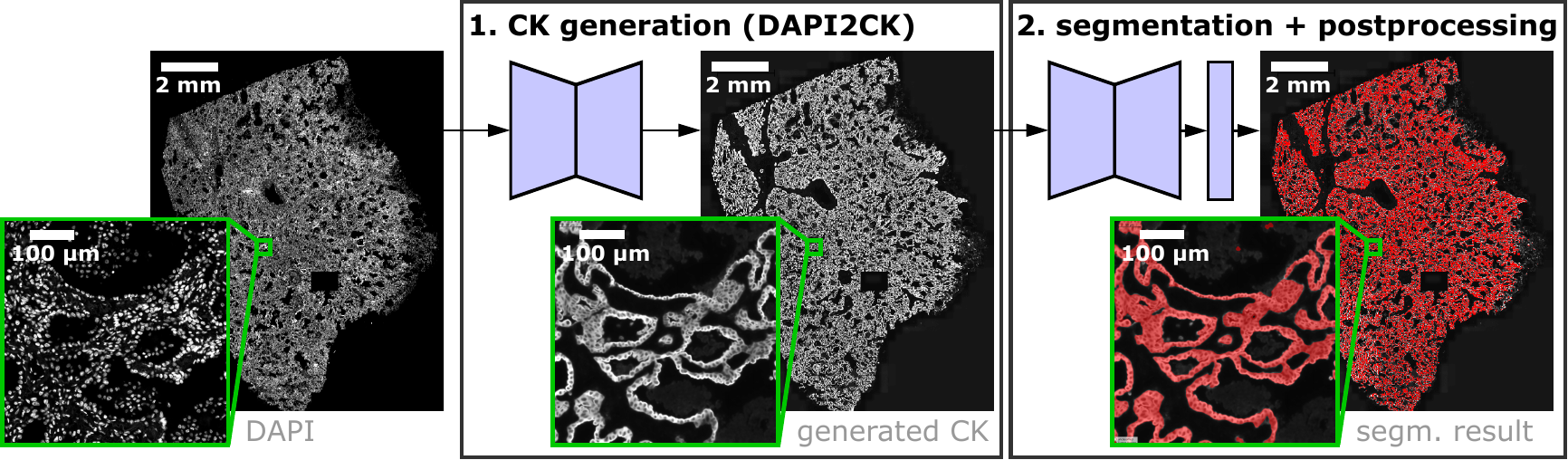}}
\end{figure}

\section{Methods}

Fig. 1 shows our approach to generate synthetic CK via \textsc{dapi2ck} and subsequently segment epithelial regions.
The \textsc{dapi2ck} generator is trained to synthesize the CK channel from the DAPI channel by training adversarially against a PatchGAN discriminator, while constraining the output image with the ground truth via an L1 loss \cite{Isola2016}. For training, patches sampled from 400 mIF whole slide images (WSIs) were used. 
To segment epithelium based on the CK channel (synthetic or real), we trained a convolutional neural network for semantic segmentation using expert annotations. The model was trained on a set of $\SI{24}{k}$ patches of the real CK channel and corresponding masks derived from epithelium annotations in 28 NSCLC images.
For both models, patches of $\num{256} \times \SI{256}{\px}$ at $\SI{0.5}{\mucron}/\si{\px}$ resolution were used for training and sliding window prediction.

\section{Results}

As shown in Fig. 2 (left panel), the \textsc{dapi2ck} output closely resembles stained CK for different tumor architectures. To evaluate the segmentations, we compared them to expert annotations within 21 fields of view (FOVs) across 5 NSCLC cases. As shown in Table \ref{tab_1}, segmentation on synthetic CK is slightly more accurate than on real CK staining. This is mostly due to the more homogenous staining pattern from \textsc{dapi2ck} compared to real staining, which requires less robust segmentation models.

\begin{table}[htbp]
\floatconts
  {tab_1}%
  {\caption{Evaluation of segmentation results.}}%
  {\begin{tabular}{rccc}
  \bfseries Test & \bfseries F1-score & \bfseries Precision & \bfseries Sensitivity\\
  stained CK vs. annotations in FOVs & 0.82 & 0.87 & 0.77\\
  \textsc{dapi2ck} vs. annotations in FOVs & 0.86 & 0.79 & 0.95\\
  \textsc{dapi2ck} vs. stained CK on full slides & 0.75 & 0.65 & 0.88
  \end{tabular}}
\end{table}

\begin{figure}[htbp]
\floatconts
  {fig_2}
  {\caption{Left panel: Comparison of synthetic and stained CK expression and corresponding segmentation results for different tumor architectures (A, B, C). Right panel: Special scenarios in which \textsc{dapi2ck} results differ strongly from stained CK.}}
  {\includegraphics[width=1\linewidth]{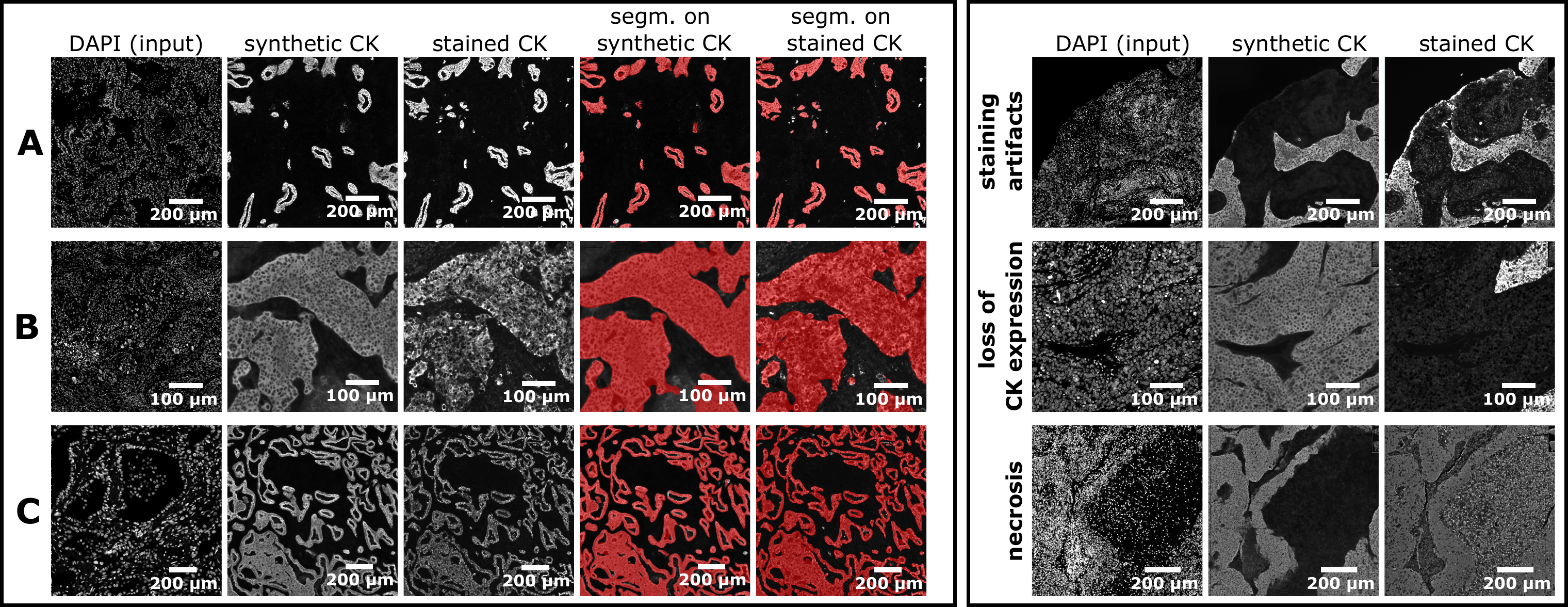}}
\end{figure}

We also compared the segmentations across 10 WSIs of NSCLC resections. With an F1-score of 0.75, segmentation on synthetic CK and stained CK largely match. However, for specific regions we see clear differences between both signals. Fig. 2 (right panel) shows three main drivers of such deviations: 1) \textsc{dapi2ck} does not suffer from staining artifacts such as unspecific staining. 2) \textsc{dapi2ck} generates staining also in epithelial regions showing no or weak CK staining (due to loss of expression or staining artifacts). 3) Necrotic areas often show CK signal, which is less pronounced in synthetic CK.

\section{Discussion}

We propose \textsc{dapi2ck} as a novel approach to derive synthetic CK staining from the DAPI channel of mIF stained pathology slides. \textsc{dapi2ck} results closely resemble real staining while artifacts from the staining process are avoided. Despite the advantages, \textsc{dapi2ck} depends on the quality of the DAPI staining. Hence, artifacts in the DAPI channel may lead to undesired effects in CK prediction. In our assay however, we found the DAPI channel to be more robust than the CK channel.
Staining slides with DAPI and a CK marker, the paired data to train \textsc{dapi2ck} is easy to generate. The approach can serve as a blueprint to leverage morphological information from one IF channel to derive expression of other markers. Our approach hence can be utilized to significantly increase the value of given mIF panels for biomarker discovery promoting patient selection for immunotherapies.

\midlacknowledgments{We thank the Prod. Informatics and mIF teams of AstraZeneca for expert support throughout the study. Results are based on the AstraZeneca Tumor \& Immune Cell Atlas (TICA).}

\bibliography{bibliography}

\end{document}